# Absence of a Band Gap at the Interface of a Metal and Highly Doped Monolayer MoS$_2$

Alexander Kerelsky[1], Ankur Nipane[2], Drew Edelberg[1], Dennis Wang[1,3], Xiaodong Zhou[1], Abdollah Motmaendadgar[1,4], Hui Gao[5,6], Saien Xie[6,7], Kibum Kang[5,6], Jiwoong Park[5,6], James Teherani[2], Abhay Pasupathy[1]

[1] Department of Physics, Columbia University, New York, NY 10027

[2] Department of Electrical Engineering, Columbia University, New York, NY 10027

[3] Department of Applied Physics and Mathematics, Columbia University, New York, NY, 10027

[4] Department of Mechanical Engineering, Columbia University, New York, NY, 10027

[5] Department of Chemistry and Chemical Biology, Cornell University, Ithaca, NY, 14853

[6] Department of Chemistry, Institute for Molecular Engineering, and Frank Institute, University of Chicago, Chicago, IL 60637

[7] School of Applied and Engineering Physics, Cornell University, Ithaca, NY, 14853

**ABSTRACT:** High quality electrical contact to semiconducting transition metal dichalcogenides (TMDCs) such as MoS$_2$ is key to unlocking their unique electronic and optoelectronic properties for fundamental research and device applications. Despite extensive experimental and theoretical efforts reliable ohmic contact to doped TMDCs remains elusive and would benefit from a better understanding of the underlying physics of the metal-TMDC interface. Here we present measurements of the atomic-scale energy band diagram of junctions between various metals and heavily doped monolayer MoS$_2$ using ultra-high vacuum scanning tunneling microscopy (UHV-STM). Our measurements reveal that the electronic properties of these junctions are dominated by 2D metal induced gap states (MIGS). These MIGS are characterized by a spatially growing measured gap in the local density of states (L-DOS) of the MoS$_2$ within 2 nm of the metal-semiconductor interface. Their decay lengths extend from a minimum of ~0.55 nm near mid gap to as long as 2 nm near the band edges and are nearly identical for Au, Pd and graphite contacts, indicating that it is a universal property of the monolayer semiconductor. Our findings indicate that even in heavily doped semiconductors, the presence of MIGS sets the ultimate limit for electrical contact.

Since the onset of mechanical exfoliation as a means to isolate thin layers of van der Waals materials, a wide array of research has been conducted on characterization, synthesis and device applications. In particular, extensive efforts have been directed towards TMDCs due to their electronic[1,2] and optoelectronic[3-6] properties. Low-resistance ohmic contacts are critical for investigating and utilizing these material properties. Ohmic contacts enable ambipolar conduction, enable high 'on' current[7,8], and allow efficient extraction of photo-response in optoelectronic devices[3,9]. A number of methods for achieving low-resistance contacts have been employed in past[10]: optimizing contact geometry (top/edge contacts)[11], optimizing contact material[12,13,14], doping underneath contacts[15,16], gating contacts[13], phase engineering[17], insertion of tunnel barriers between the metal and semiconductor[18], etc. However, despite extensive experimental and theoretical[19,20] efforts, reliable high-quality contact to these materials still remains elusive and efforts towards it are especially hindered by a lack of understanding[21] of the atomic-scale physics at the metal-TMDC interfaces.

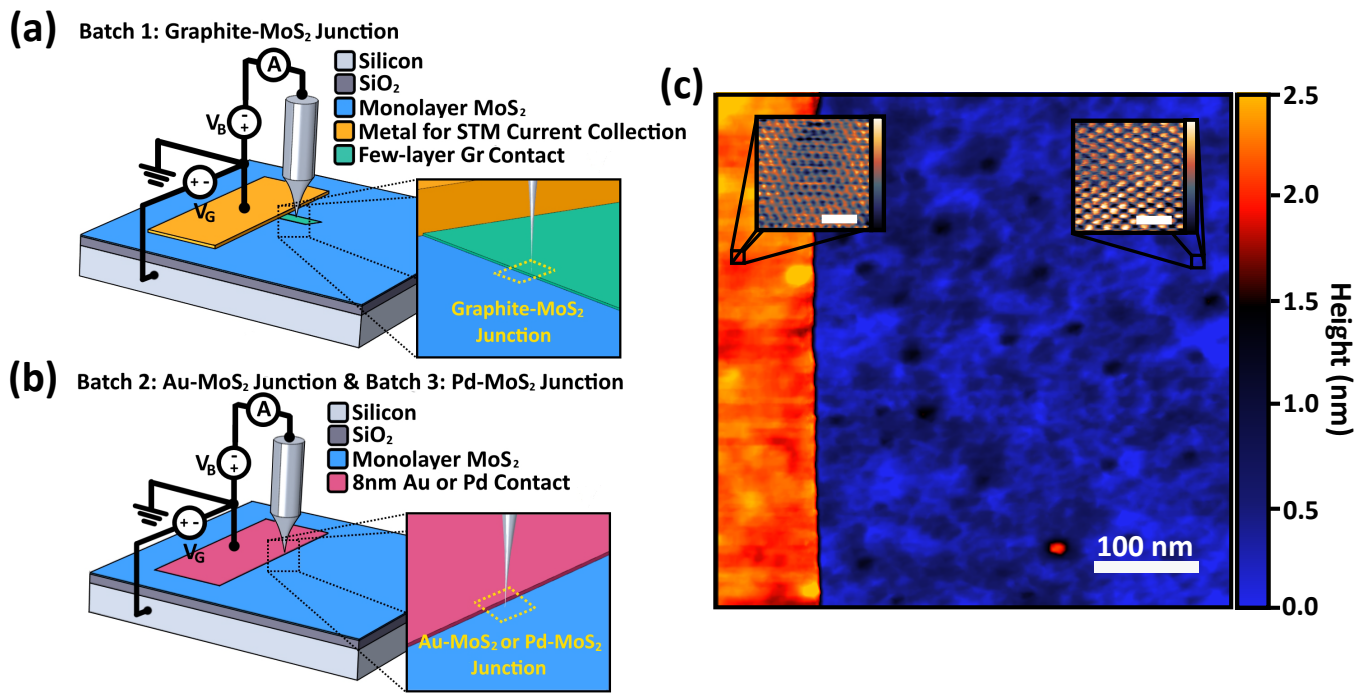

**Figure 1. (a)** Schematic of the device and experimental setup for graphite-MoS$_2$ junction. $V_B$ is the sample-probe bias voltage and $V_G$ is the gate voltage. Topographies and line spectroscopies are taken across the edge of the few-layer graphite contact, as highlighted by the yellow dotted box in the zoom-in inset. **(b)** Schematic of the device and experimental setup for Au-MoS2 or Pd-MoS2 junctions made by evaporating thin sharp Au or Pd contacts through a shadow mask 5 μm from the monolayer MoS$_2$ surface. Topographies and line spectroscopies are taken across the edge of the contact as highlighted by the yellow dotted box in the zoom-in inset. **(c)** A representative STM topographic image of the edge of a graphite electrode atop the monolayer MoS$_2$ film. The image shows the sharpness of the junction as well as the uniformity of both the MoS$_2$ and graphite adhering to the SiO$_2$ substrate. The insets show atomic resolution topographies for each material, which have been used to confirm the lattice constants of each respective material. The white scale bar in the insets is 1 nm and the intensity bars represent 0 to 150 pm for graphite and 0 to 200 pm for MoS$_2$. STM topography set points are 1 V, 300 pA for the large area junction, 3 V, 100 pA for the graphite inset and 4 V, 500 pA for the MoS$_2$ inset.

How the properties of top contact metals correlate with Schottky barrier height, contact resistance, and band alignments is an essential component that is not well understood[22,23]. Most previous studies of contact properties have been performed by transport and optical techniques. While both of these can shed light on overall properties of the contacts such as the contact resistance and the difference in work function, they do not offer the spatial resolution that is key to understanding the precise band alignment as well as the lateral properties at the contact edge.

Ultra-high vacuum scanning tunneling microscopy (UHV-STM) – the probe used in this study – provides the atomic-scale resolution necessary to investigate the lateral properties and precise band alignment but is experimentally more challenging for a number of reasons. Ultra-clean, conducting samples are necessary[24,25], which has been hindered by residue from photoresist and standard polymer transfer techniques, as well as the difficulty of performing STM on an insulating substrate. A sharp contact edge is also imperative for an abrupt metal-semiconductor junction. Furthermore, optical resolution limitations resulting from the large optical working distance from outside a UHV chamber to samples within the UHV chamber also make it difficult to approach STM probes to small-area TMDC samples.

In this work, we fabricate <10-nm-thick top contacts with nanometer-scale edge sharpness atop high-quality, large-area, heavily n-type monolayer MoS$_2$ films allowing UHV-STM and scanning tunneling spectroscopy (STS) atomic-scale characterization of the interface. The large n-type carrier concentration is used in order to minimize Schottky barrier effects and examine contacts near the ohmic regime. We also investigate the impact of different metal properties by studying three different types of metal-MoS$_2$ top contact junctions. Graphite and gold (Au) were chosen as metals of interest due to their use as reasonably low-resistance top contacts to $MoS_2$[26]. Palladium (Pd) was also chosen due to its high work function and previous works achieving p-type contact to $MoS_2$ using Pd[27].

Monolayer $MoS_2$ films were grown directly on $Si/SiO_2$ substrates with nearly uniform growth across 4-inch wafers[28]. More than 95% of the film area consisted of monolayer $MoS_2$ with an occasional patch of bilayer or trilayer $MoS_2$ (such patches were avoided in this study). Optical absorption, photoluminescence (PL), and Raman spectroscopy were used for preliminary sample quality characterization. Monolayer films showed a PL peak at 1.87 eV, in confirmation with previous PL on high quality exfoliated and CVD samples in other works. After preliminary film quality characterization, wafers were cleaved into 3 mm × 10 mm pieces and split into three batches for fabrication with different contact metals.

In Batch 1, a graphite-$MoS_2$ top contact junction was created by depositing a thin exfoliated flake of graphite (about 2-nm-thick) — providing a naturally sharp contact edge — onto monolayer $MoS_2$ using the standard polymer dry transfer technique. To enable collection of the STM tunneling current, contact to the graphite and $MoS_2$ was formed by Au evaporation through a shadow mask to preserve sample cleanliness (see Figure 1(a)). Note that for Batch 1, in contrast to Batches 2 and 3, the evaporated metal is merely for collecting tunneling current because the graphite-$MoS_2$ junction of interest is atop an insulating substrate of $SiO_2$. Figure 1(a) shows a schematic of the fabricated structure as well as the electrical connections for the STM experiment. For Batches 2 and 3, 8 nm of Au or Pd, respectively, was directly deposited atop the $MoS_2$ film (without exfoliated graphite), as contacts of interest for metal-$MoS_2$ junction characterization. The metals were evaporated through a shadow mask that was designed and positioned to lie only ~5 µm above the $MoS_2$ film, producing sharp Au-$MoS_2$ and Pd-$MoS_2$ junctions, while preserving the cleanliness of the samples. Figure 1(b) is a schematic of the Au-$MoS_2$ and Pd-$MoS_2$ sample batches.

The samples were then loaded into a UHV scanning tunneling microscope and annealed at 100 C for a minimum of two hours, to minimize surface contaminants. During UHV-STM measurements, the samples (metal contacts and $MoS_2$ films) were grounded and the probe was biased to establish tunneling current. All measurements were taken at room temperature since contact resistance becomes very large at cryogenic temperatures preventing accurate measurement of STM tunneling current. To map the spatial local density of states (L-DOS) of the junctions, STS dI/dVs were taken at 204 equally spaced points across each metal-$MoS_2$ junction (341 for 2.5 µm line profile). For each dI/dV, a sample-probe voltage difference of 2 V and initial set point of 500 pA for graphite (100 pA for Au and Pd) was used to adjust the sample-probe distance, after which the feedback was frozen and sample probe voltage was ramped from 2 V to -2 V while measuring tunneling current. dI/dVs, proportional to the L-DOS, were calculated by taking numerical derivatives of the STS tunneling current profiles. Note that dI/dV only provides a value proportional to the L-DOS rather than the actual magnitude. Thus, the method can be used to determine band gaps and the L-DOS functions up to a multiplicative constant. This constant can be treated as nearly uniform within the individual materials due to the uniformity of the L-DOS within each material at the voltage set-point of 2 V (which determines the constant), thus the spatially varying L-DOS within each individual material can be compared; however, because of the difference in L-DOS at 2 V between different materials, only the L-DOS shapes can be compared between different materials.

Figure 1(c) shows an STM topographic image at the edge of the graphite contact atop the $MoS_2$ film. The image reveals the sharpness of the graphite-$MoS_2$ junction, as well as the uniformity of the monolayer $MoS_2$ film and the graphite top contact. The two materials adhere to the natural roughness of the $SiO_2$ substrate. The $MoS_2$ roughness is comparable to the roughness of single-layer graphene on $SiO_2$, while the graphite roughness is significantly less due to the stiffness of a larger number of stacked layers. The sharp, abrupt boundary at the graphite-$MoS_2$ interface, as well as the cleanliness of the surfaces of both materials, are ideal for spatial L-DOS characterization. Topography of the Au-$MoS_2$ and Pd-$MoS_2$ junctions are provided in the supplemental information. The evaporated Au and Pd top contacts are inherently a coalescence of grains from the evaporation deposition. Thus, they do not have perfectly straight edges like exfoliated graphite. Given this limitation, the closely placed shadow masks through which the Au and Pd were evaporated ensure abruptly ending edges defined by the local grains at the edges, allowing L-DOS characterization across the Au-$MoS_2$ and Pd-$MoS_2$ junctions as well.

We first investigate the material properties of monolayer $MoS_2$ and graphite at large length scales by analyzing a 2.5 µm dI/dV line profile across the graphite-$MoS_2$ junction (Figure 2(a)). The figure shows a heat map of the dI/dVs as a function of position and sample-probe voltage difference (all normalized to the same value at 2 V in order to put all spectra on the same colorscale), spatially aligned with a concurrently-measured sample-probe displacement below the heat map. This long dI/dV spatial profile provides the L-DOS of graphite and $MoS_2$ far away from the junction, which allows prediction of the graphite-$MoS_2$ band alignment and verifies the spatial uniformity of the L-

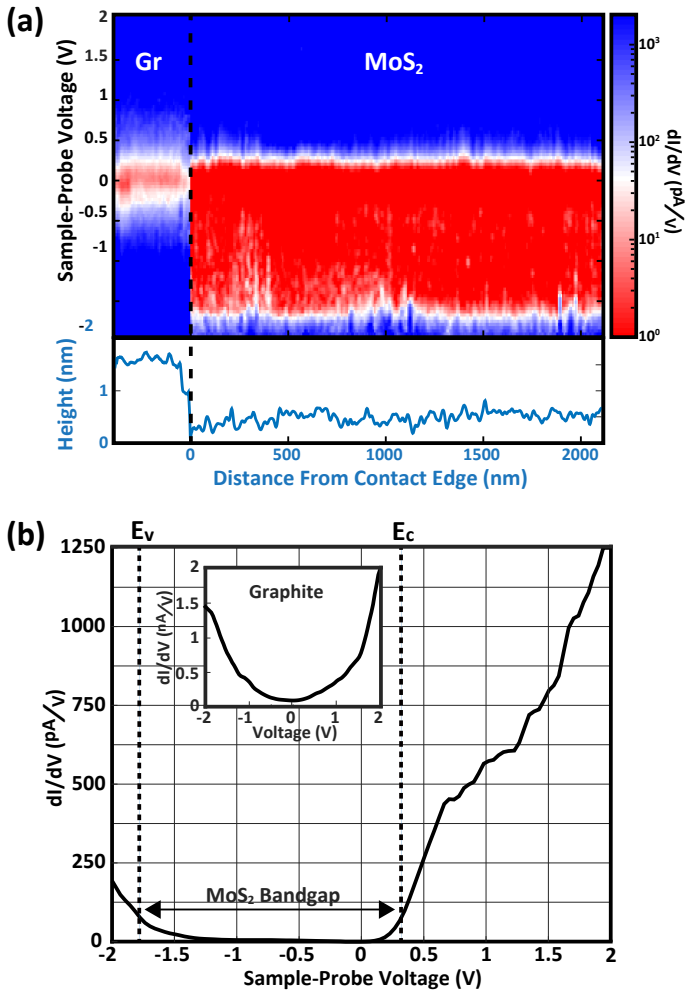

**Figure 2. (a)** Heat map of a long (~2.5 μm) dI/dV line profile taken across the edge of the graphite electrode, consisting of 341 equally spaced points. Below the heat map is the simultaneously measured STM topography profile revealing the precise location of the graphite-MoS$_2$ step edge. The length scales have been offset such that 0 is the junction location. The heat map reveals an abrupt change in the dI/dV, and hence the L-DOS, profile on this 2.5 μm-sized scale, after which the MoS$_2$ band gap emerges, seen by the low dI/dV intensity region (red). The heat map also demonstrates that the dI/dV of the separate materials are approximately uniform on this scale. **(b)** A spatially averaged dI/dV of the MoS$_2$ and the graphite (inset), representing the DOS far from the junction. The graphite DOS exhibits no band gap, whereas a band gap of ~2.1 eV is observed for MoS$_2$, similar to the quasiparticle band gap observed in previous STM studies of monolayer MoS$_2$. The asymmetry in the position of the valence and conduction band edges of the MoS$_2$ (with respect to 0 V) implies that our MoS$_2$ film is n-type.

DOS at large length scales. It is clear from the dI/dV line profile that there is an abrupt change in the dI/dVs at the graphite-MoS$_2$ junction, signifying a change in the L-DOS, after which there is long-range uniformity of L-DOS of both graphite and MoS$_2$ individually. Small fluctuations in the L-DOS can be attributed to defects in the materials.

On the MoS$_2$ side of the dI/dV heat map in Figure 2(a), the region of nearly zero L-DOS (red color) indicates the band gap of the material, clearly not present on the graphite side. To analyze the details of the dI/dV profile of each material far from the junction, Figure 2(b) displays a spatially-averaged dI/dV curve of the monolayer MoS$_2$, and in the inset, the graphite. These averaged dI/dVs can be interpreted as the DOS of the graphite and MoS$_2$ when not modified by junction physics. The graphite DOS exhibits metallic properties, as there exists no region with zero DOS as a function of sample-probe voltage, hence there is no band gap. All three contact metals (Pd, Au, graphite) studied show a similar metallic DOS, consistent over many different samples. The MoS$_2$ DOS shows a considerable band gap, as the DOS plunges to nearly zero from about -1.75 V to about +0.35 V alluding to a quasiparticle gap size of about 2.10 eV, similar to that of previous reports on MoS$_2$[24, 25]. In this conversion of quasiparticle band gap energy size in eV from dI/dV sample-probe voltage thresholds, it has been calculated that tip induced band bending and image charge potential are nearly equal and opposite and thus allow direct conversion of the observed gap in the dI/dV spectrum to quasiparticle band gap[25]. It is clear from the dI/dV that the conduction band is closer to the Fermi level (represented by zero sample-probe voltage) than the valence band implying that the MoS$_2$ film is n-type. The n-type behavior is in agreement with typical CVD MoS$_2$ samples that have been studied both by transport and STM measurements[29]. Further, we observe band-tails (characterized by ill-defined band edges beyond what is expected by the Fermi-Dirac broadening) signifying large effective sheet charge concentrations[30] which we use to study contacts near the ohmic regime. It is tempting to determine the precise carrier concentration based on the apparent conduction band edge location with respect to the Fermi level, however, the presence of an unknown amount of contact doping caused by the probe makes this inaccurate.

Next, we investigate the local properties of graphite-MoS$_2$ junction by analyzing a short spatial dI/dV profile across the junction. Figure 3(a) shows a heat map of the dI/dV calculated from a ~9-nm STS line scan of across the junction with the concurrently measured sample-probe displacement. We define the precise junction position where the topographic height reaches that of the MoS$_2$ layer. Areas to the left of the junction are the metal, while areas to the right are the semiconductor. It is important to note for the analysis following that the topography clearly separates metal and semiconductor regions from each other, i.e., there are no effects due to the tip radius in these measurements that are commonly seen in AFM measurements. While the probe is above the graphite, the dI/dV heat map shows that the electronic spectra remain metallic, similar to the graphite region observed in the long dI/dV line profile in Figure 2(a). Once the probe crosses the junction, a finite gap begins to grow over the next 2 nm. This can be seen as both the valence and conduction band edges begin near 0 V and gradually shift toward their long range MoS$_2$ film values, observed earlier in Figure 2(a). No clear band-bending and depletion region is observed in our measurements. Heat maps of Au-MoS$_2$ and Pd-MoS$_2$ junctions in Figure 3(b) and (c) reveal the same behavior. To ensure that this effect is not a heat map/plotting artifact, Figure 3(d) shows several individual dI/dVs at different spatial positions in the region of the evolution from the graphite-MoS$_2$ interface. Comparing the dI/dVs, we see a gradual formation of the valence band edge based on the negative voltage tails, which begin metallic and gradually become more and more flat, approaching the profile seen earlier in Figure 2(b). We can also see the conduction band edge forming, although masked by its proximity to the Fermi level. Thus, the L-DOS profiles show that the dI/dV gap grows with increasing distance from the edge of the graphite-MoS$_2$ junction. The same effect is observed in Au-MoS$_2$ and Pd-MoS$_2$ on similar length scales, as can be verified by Figures 3(b) and (c).

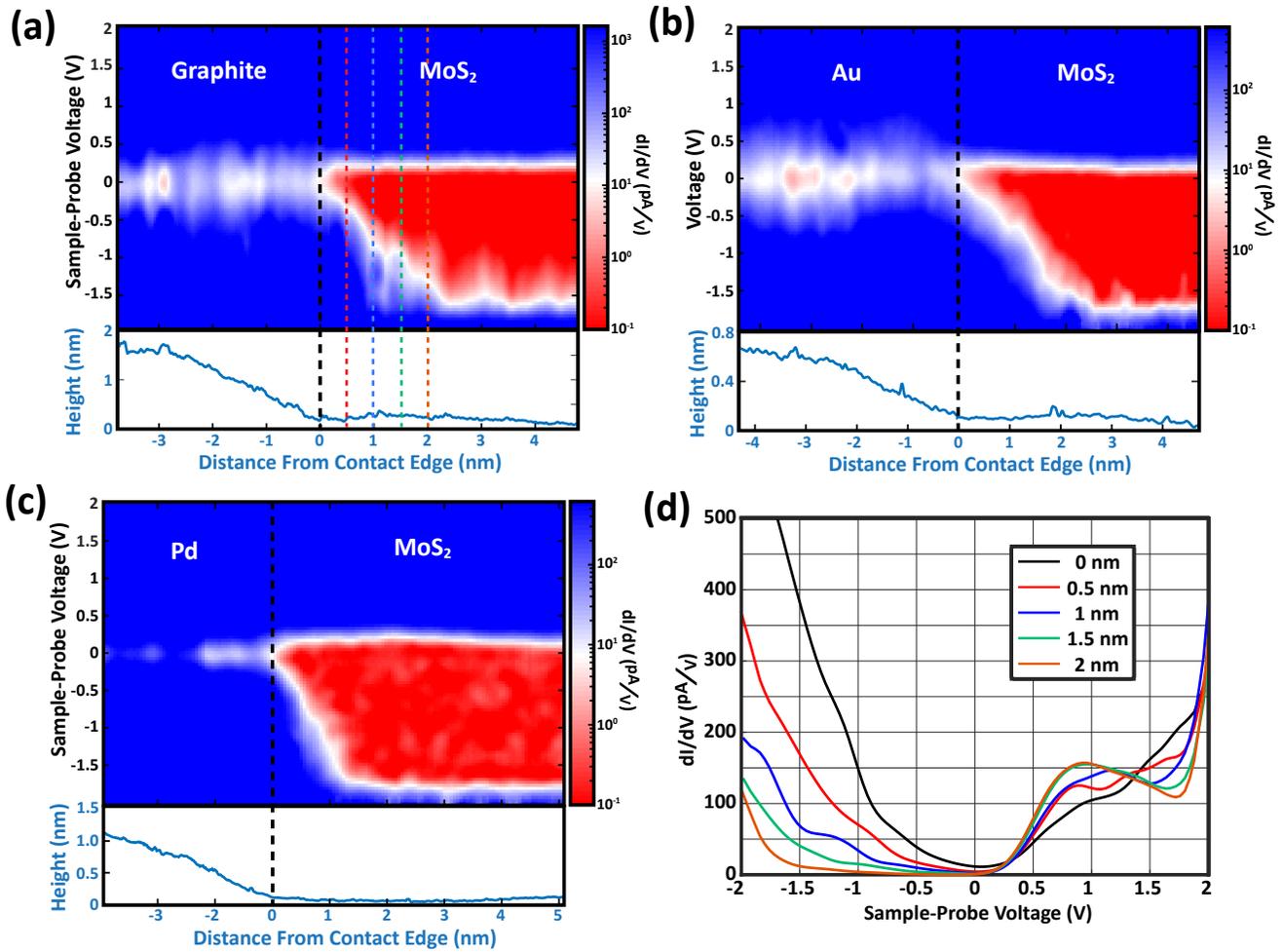

**Figure 3. (a-c)** Heat maps of *short* (~9-nm) dI/dV line profiles taken across the **(a)** graphite-MoS$_2$, **(b)** Au-MoS$_2$, and **(c)** Pd-MoS$_2$ junctions consisting of 204 equally spaced points. Below the heat map is a simultaneously measured STM topography profile revealing the location of the contact edge. The heat map reveals a metallic L-DOS profile up until the edge of the electrode. Immediately to the right of the interface into the MoS$_2$ side, there is still no observed L-DOS gap. A L-DOS gap emerges at a finite distance into the MoS$_2$ side and grows for about 2 nm before reaching the far from junction magnitude. **(d)** Individual dI/dV line spectra taken at 0, 0.5, 1, 1.5, and 2 nm from the junction, as shown by the dashed lines in Figure 3(a). The individual line spectra confirm the apparent evolution seen in the heat map, as the conduction and valence band edges are gradually defined with distance from the junction, reaching the far from junction, full MoS$_2$ band gap several nanometers from the junction.

We find that the observations are best explained by the continuum of metal-induced gap states (MIGS)[31, 32]. Heine showed that the presence of metal at the interface creates evanescent states inside the semiconductor band gap which decay exponentially from the junction. These tailing states could be visualized as virtual gap states of the complex band structure of the semiconductor as shown in Figure 4(a). For states that lie within the band gap, only the imaginary part of wavevector k exists, resulting in exponential decay characterized by the decay length, $\delta = \frac{1}{q}$, where $q = ik$ and k is the standard plane wave vector. Mönch developed a one-dimensional virtual gap state model to quantify this virtual gap state decay length (Supplementary section)[33]. The model reveals that as a function of energy the decay lengths diverge at the band edges, whereas the decay lengths are minimum at the charge neutrality level (CNL), as plotted in Figure 4(b). The CNL is at the center of the band gap in this model, but shifts when effective electron and hole masses have any mismatch, which is the case of $MoS_2$. The minimum wave function decay length in this model is $\delta_{min} = \frac{2\pi\hbar^2}{m_e a E_g}$, where $\hbar$ is the Planck constant, $m_e$ is the free electron mass, $a$ is the lattice constant, and $E_g$ is the semiconductor band gap.

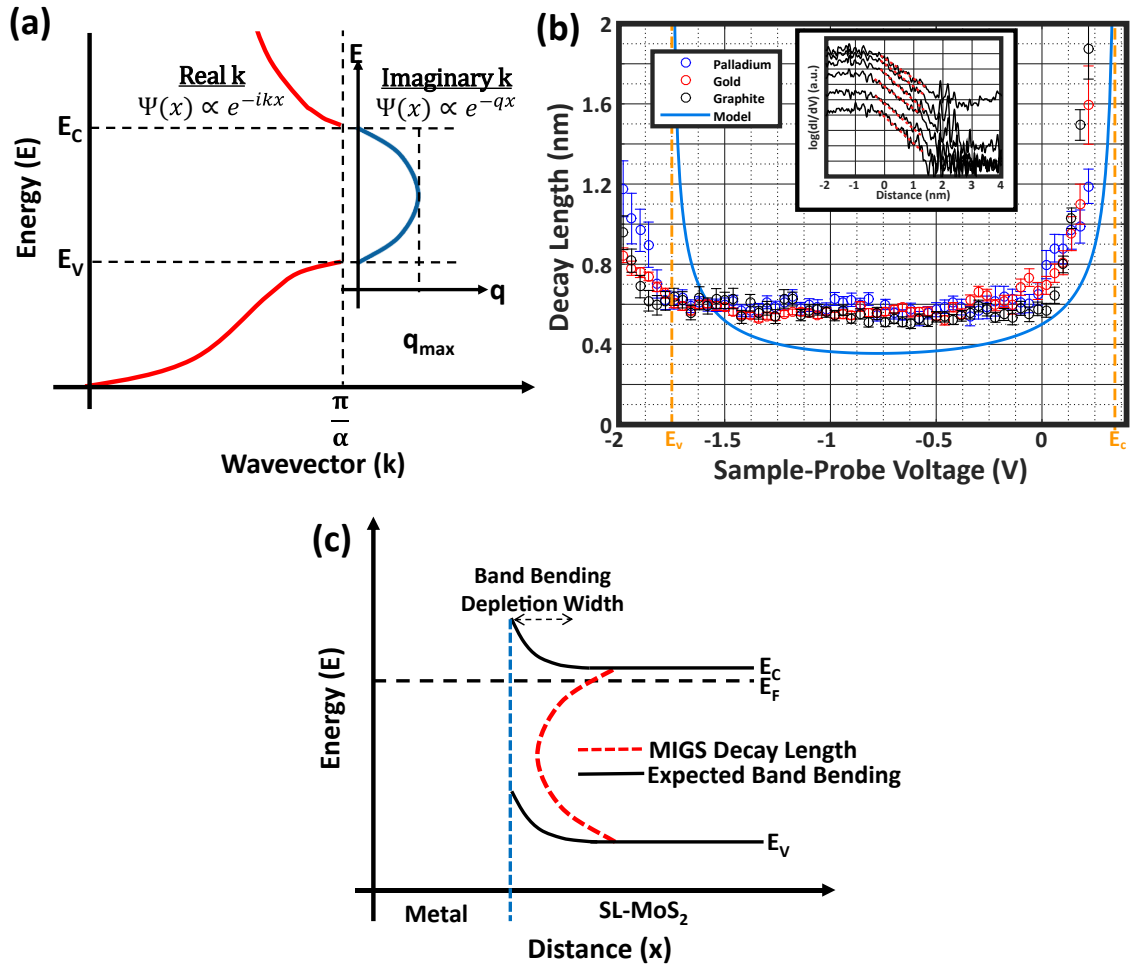

**Figure 4. (a)** The band structure calculated in the simple model used to predict the MIGS decay lengths. The red curves show the real wave vectors, k, obtained for electrons in a 1D periodic potential, producing a band gap at the Brillioun zone boundary. The blue curve demonstrates the virtual states for imaginary k values ($q = ik$), within the band gap of the semiconductor, which characterize the MIGS decay lengths. In this simple model, the maximum q and minimum decay length (where $\delta = 1/q$) are at mid gap while, at the band edges, q approaches 0 which leads to divergent decay lengths. **(b)** Plot of the experimental MIGS decay lengths into $MoS_2$ calculated from linear fits at each energy for Pd-$MoS_2$, Au-$MoS_2$ and graphite-$MoS_2$ junctions. The theoretical curve, derived from a 1D periodic potential model, is also plotted for reference. The experimental minimum decay length is about 0.55 nm at an energy slightly shifted towards the conduction band minimum from mid-gap due to the CNL shift from mid-gap in $MoS_2$. The inset shows several example spatial L-DOS profiles at fixed voltages (from top to bottom, -1.98, -1.62, -1.26, -0.90, -0.54, -0.18 V), plotted on log scale, confirming the linearity, hence the exponential behavior of the MIGS. **(c)** Energy band diagram for an ideal metal-semiconductor Schottky barrier, as well as the MIGS decay lengths scaled longer than the characteristic band bending depletion width demonstrating why experimental results do not exhibit a Schottky barrier.

The decay length of the L-DOS is half the value of the wave function decay length, as the charge density is proportional to the wave function squared. Using the experimental values of in-plane lattice constant $a = 0.32$ nm and single particle band gap $E_g = 2.1$ eV for monolayer MoS$_2$, the theoretical minimum decay length for metal-MoS$_2$ interfaces is 0.36 nm.

To confirm the presence of MIGS, we investigate the spatial evolution of the MoS$_2$ L-DOS within the band gap, starting at the edge of the contact. The inset of Figure 4(b) shows the logarithm of dI/dV intensity at a series of fixed energies within the band gap, as a function of distance from the graphite-MoS$_2$ interface. The linear decrease in log(dI/dV) with distance from the junction signifies the exponential decay of the MIGS, in confirmation with theory. Linear regression fits on position versus log(dI/dV) at each voltage within the band gap, allow determination of the experimentally observed decay length of the MIGS. Figure 4(b) shows the experimental decay lengths as a function of energy for the three different metal-MoS$_2$ junctions, Au-MoS$_2$, Pd-MoS$_2$ and Graphite-MoS$_2$, as well as the decay lengths predicted by the previously mentioned theoretical model. The MIGS decay lengths for the three contact metals are almost identical, as the decay length is determined solely by semiconductor parameters, as in Mönch's model (supplementary information). The minimum decay length of $0.55 \pm 0.10$ nm occurs around -0.6 V, slightly shifted towards the conduction band edge from mid-gap. This experimental decay length is quite close to the theoretical expectation of 0.36 nm, considering the simplicity of the model which only accounted for the effect of the periodic potential of the lattice. The shift of the minimum decay length in energy from mid-gap is due to the mismatch in the effective electron and hole masses in MoS$_2$[34], which shifts the CNL from mid-gap. It has been found that the CNL in monolayer MoS$_2$ is shifted towards the conduction band[22], the same direction as the shift in minimum decay length in our experimental results. At the conduction band edge, we observe the expected divergence in the MIGS decay lengths, in concurrence with the theoretical curve. Approaching the valence band edge, there is a gradual increase in decay length, although the MIGS persist into the valence band before diverging. This continuation of MIGS into the valence band, in disagreement with models, has been previously observed[35]. The inability of the simple model to capture this effect is likely because the model fails to account for the precise nature of the valence band and the localization of carriers near the valence band edge, which is affected by defects and film quality. MIGS have been experimentally observed in 1D[35] and 3D[36] systems, but to our knowledge, this is the first experimental confirmation of MIGS in 2D materials.

Another interesting aspect of our findings is the lack of the inherent Schottky barrier and depletion width formed at the metal-MoS$_2$ junctions due to work-function mismatch. Although the large carrier concentration was expected to minimize the Schottky barrier, the complete absence is surprising. Schottky barriers are normally characterized by spatial band-bending inside the semiconductor which indicates the depletion region inherent due to Fermi level misalignment, as shown in Figure 4(c). This is clearly not observed in our experimental dI/dV spatial line profiles. Thus, there is no Schottky barrier in these contacts and MIGs govern the metal-semiconductor transition and set the ultimate limit for electrical contact in heavily doped monolayer semiconductors. We attribute the lack of a Schottky barrier to a shorter depletion width than the characteristic MIGS decay length as indicated in Figure 4(c). The depletion width is inhibited by the effective sheet concentration of our MoS$_2$ film which we can estimate by employing an analytic model for 2D depletion width ($w_{2D}$) given by Gurugubelli et. al., where $w_{2D} = \frac{4\epsilon_{eff}(\phi_{bi}-V)}{\pi q N_{2D}}$ [37,38]. Here, $\epsilon_{eff}$ is effective dielectric constant, $\phi_{bi}$ is the built-in potential, $V$ is the applied bias, $q$ is the elementary charge and $N_{2D}$ is the effective sheet carrier concentration. As opposed to the standard 3D model, the 2D model is suited for systems with 2D materials such as our metal-MoS$_2$ junctions as it considers the role of the significant out-of-plane electric field which is absent in 3D model[39]. Based on the 2D depletion width model, we can assume an effective sheet carrier concentration of at least $10^{13}$ $Carriers/cm^2$ in order to yield a depletion width shorter than the MIGS spatial extent, a reasonable value for degenerate n-type MoS$_2$ film[16]. We also find that the application of a back gate voltage from -20V to 80V (corresponding to a sheet carrier density of $5 \times 10^{12}$ $Carriers/cm^2$) does not significantly shift the dI/dV spectrum, further verifying this degenerate carrier concentration. This agrees with our previously mentioned findings that the MoS$_2$ film has a high carrier concentration due to the presence of band tails in the DOS. Future investigations of MoS$_2$ and other TMDC samples that are more intrinsic will further the understanding of contacts by allowing direct observations of the Schottky barrier, depletion width, and MIGS, which can be used to compare metals, study gate dependence, and investigate novel methods to avoid Fermi level pinning (which has in the past been attributed to MIGS)[40].

In conclusion, we investigated the metal-MoS$_2$ junction of three different top contacts—graphite, Au and Pd—on heavily n-type monolayer MOCVD grown MoS$_2$ using UHV-STM and STS. By fabricating clean nanoscale sharp contact edges on large area MoS$_2$ films atop SiO$_2$, we have provided sub-nanometer-scale spatial spectroscopic characterization of the evolution of the dI/dV, proportional to the L-DOS, of the $MoS_2$ in the nanoscale vicinity of all three metal-$MoS_2$ junctions. dI/dV line profiles across the junctions reveal a gradually growing L-DOS gap in the $MoS_2$, originating at the metal-$MoS_2$ junction. The effect is attributed to MIGS originating from the contact metals, decaying into the $MoS_2$. By analyzing the energy dependence of the decay length of the MIGS, we observe the decay length can vary from a minimum of $0.55 \pm 0.10$ nm to greater than 1 nm near the band edges, in good agreement with theory. The MIGS decay length is shown to be independent of the contact metal and solely determined by the parameters of the semiconductor, as predicted in theory. Also, we show that in these contacts, there is a lack of a Schottky barrier, concealed due to a shorter depletion width than the observed MIGS decay length indicating that MIGS set the ultimate limit for highly doped monolayer semiconductor electrical contact.

## SUPPORTING INFORMATION

Model calculation of metal induced gap states, STM topography of Au-$MoS_2$ and Pd-$MoS_2$ junctions are attached.

## ACKNOWLEDGEMENTS

We thank Jerry Tersoff, Christopher Gutiérrez and Elton Santos for discussions. This work is supported by the National Science Foundation (NSF) Materials Research Science and Engineering Center at Columbia University (grant DMR 1420634) and by the Air Force Office of Scientific Research (grant number FA9550-16-1-0601, A.K. and FA9550-16-1-0031, J.P.). Material synthesis was supported by the NSF through the Platform for the Accelerated Realization, Analysis, and Discovery of Interface Materials (PARADIM; DMR-1539918) and the Cornell Center for Materials Research (NSF DMR-1120296).

## CORRESPONDING AUTHOR

Abhay N. Pasupathy: apn2108@columbia.edu

# Supporting Information

# Absence of a Band Gap at the Interface of a Metal and Highly Doped Monolayer MoS$_2$

Alexander Kerelsky[1], Ankur Nipane[2], Drew Edelberg[1], Dennis Wang[1,3], Xiaodong Zhou[1], Abdollah Motmaendadgar[1,4], Hui Gao[5,6], Saien Xie[6,7], Kibum Kang[5,6], Jiwoong Park[5,6], James Teherani[2], Abhay Pasupathy[1]

[1] Department of Physics, Columbia University, New York, NY 10027

[2] Department of Electrical Engineering, Columbia University, New York, NY 10027

[3] Department of Applied Physics and Mathematics, Columbia University, New York, NY, 10027

[4] Department of Mechanical Engineering, Columbia University, New York, NY, 10027

[5] Department of Chemistry and Chemical Biology, Cornell University, Ithaca, NY, 14853

[6] Department of Chemistry, Institute for Molecular Engineering, and Frank Institute, University of Chicago, Chicago, IL 60637

[7] School of Applied and Engineering Physics, Cornell University, Ithaca, NY, 14853

**S1. Metal Induced Gap State Model**

To model metal induced gap states (MIGS), we employ a simple 1D model of an electron in a periodic potential

$$V(x) = 2A \cos(\frac{2\pi}{a}x) \quad (S.1)$$

Here, a is the lattice constant and A is the magnitude of the potential, later to be determined as E$_g$/2. We solve the Schrodinger Equation using a Fourier expansion of our wave function

$$\psi(x) = \sum_k C(k)e^{ikx} \quad (S.2)$$

In this equation, $k = \frac{2\pi n}{a}$ where n is an integer, preserving the boundary conditions of our lattice with lattice constant a. Plugging this into Schrodinger's equation, one obtains

$$\sum_k (\frac{\hbar^2 k^2}{2m} - E)C(k)e^{ikx} + \sum_k V(x)C(k)e^{ikx} = 0 \quad (S.3)$$

Provided that the amplitude of the potential, A, is small compared to the kinetic energy of the electrons, we can truncate this summation to two equations

$$\begin{bmatrix} \dfrac{\hbar^2 k^2}{2m} - E & A \\ A & \dfrac{\hbar^2 (k - \frac{2\pi}{a})^2}{2m} - E \end{bmatrix} \begin{bmatrix} C(k) \\ C(k - \frac{2\pi}{a}) \end{bmatrix} = 0 \tag{S.4}$$

In this simplification, we are considering the two free electron bands centered at 0 and $\frac{2\pi}{a}$, and their interaction. We look for solutions at the edge of the first Brillioun zone where the two bands intersect, $\frac{\pi}{a}$. We make a change of variables: $k = k' + \frac{\pi}{a}$, thus $k'$ is centered around $\frac{\pi}{a}$. Our matrix becomes

$$\begin{bmatrix} \dfrac{\hbar^2 (k' + \frac{\pi}{a})^2}{2m} - E & A \\ A & \dfrac{\hbar^2 (k' - \frac{\pi}{a})^2}{2m} - E \end{bmatrix} \begin{bmatrix} C(k' + \frac{\pi}{a}) \\ C(k' - \frac{\pi}{a}) \end{bmatrix} = 0 \tag{S.5}$$

As the determinant of the matrix is equal to zero, we can easily solve for $k'$. We first solve for energy

$$E(k') = \frac{\hbar^2}{2m} k'^2 + \frac{\hbar^2}{2m}\left(\frac{\pi}{a}\right)^2 \pm \sqrt{4\frac{\hbar^2}{2m}(k'^2)\frac{\hbar^2}{2m}\left(\frac{\pi}{a}\right)^2 + A^2} \tag{S.6}$$

At the edge of the first Brillioun zone, $k' = 0$, we see that the energy is

$$E(0) = \frac{\hbar^2}{2m}\left(\frac{\pi}{a}\right)^2 \pm A \tag{S.7}$$

Thus we confirm that a gap has emerged of gap size $2A = E_g$, centered at $\frac{\hbar^2}{2m}\left(\frac{\pi}{a}\right)^2$.

To determine the characteristic decay length of the virtual wave functions, we solve for $k'$ as a function of energy. Next we set $k' = iq$ to solve for the virtual states within the band gap

$$\frac{\hbar^2}{2m} k'(E)^2 = E + \frac{\hbar^2}{2m}\left(\frac{\pi}{a}\right)^2 \pm \sqrt{4\frac{\hbar^2}{2m}\left(\frac{\pi}{a}\right)^2 E + A^2} \tag{S.8}$$

$$\frac{\hbar^2}{2m} q(E)^2 = -E - \frac{\hbar^2}{2m}\left(\frac{\pi}{a}\right)^2 \mp \sqrt{4\frac{\hbar^2}{2m}\left(\frac{\pi}{a}\right)^2 E + A^2} \tag{S.9}$$

In order to have a nonnegative decay length, the nonnegative (addition sign) solution is chosen. By definition, our decay length for the virtual electron functions, $\psi = Ce^{-qx}$ are given by $\delta = \frac{1}{q}$. Thus

$$\delta = \sqrt{\frac{\frac{\hbar^2}{2m}}{-E - \frac{\hbar^2}{2m}\left(\frac{\pi}{a}\right)^2 + \sqrt{4\frac{\hbar^2}{2m}\left(\frac{\pi}{a}\right)^2 E + \frac{E_g^2}{4}}}} \quad (S.10)$$

We first minimize this function to find the minimum MIGS decay length

$$\delta_{min} = \frac{2\pi\hbar^2}{E_g m_e a} \quad (S.11)$$

This is at an energy value of

$$E(q = \delta_{min}) = \frac{\hbar^2}{2m}\left(\frac{\pi}{a}\right)^2 - \frac{\frac{1}{4}A^2}{\frac{\hbar^2}{2m}\left(\frac{\pi}{a}\right)^2} \approx \frac{\hbar^2}{2m}\left(\frac{\pi}{a}\right)^2 \quad (S.12)$$

The simplification in S.12 can be done due to our earlier assumption that the interaction potential is much smaller than the electron energies, which is given by $\frac{\hbar^2}{2m}\left(\frac{\pi}{a}\right)^2$. It is clear from equation (S.12) that the minimum decay length in Eq. (S.10) is mid-gap (which can be seen in Eq. (S.7)). The decay length of the MIGS L-DOS is half the decay length of the wave functions, $\frac{1}{2}\delta_{min}$, as the charge density is proportional to the square of the wave functions.

One can further easily confirm that decay length $\delta$ at the band edges, given by Eq. (S.7), diverge, by plugging the band edges into Eq (S.10)

$$\delta\left(E = \frac{\hbar^2}{2m}\left(\frac{\pi}{a}\right)^2 \pm \frac{E_g}{2}\right) = \frac{\frac{\hbar^2}{2m}}{\sqrt{-2\frac{\hbar^2}{2m}\left(\frac{\pi}{a}\right)^2 \mp \frac{E_g}{2} + \sqrt{\left(2\frac{\hbar^2}{2m}\left(\frac{\pi}{a}\right)^2 \pm \frac{E_g}{2}\right)^2}}} \quad (S.13)$$

Further simplification of the denominator leads to the divergent behavior.

**S2. Au-MoS₂ and Pd-MoS₂ Topographies**

Figures S1a and S1b show STM topographies of the Au-MoS$_2$ junction. Similar to the graphite-MoS$_2$ junction, the MoS$_2$ is uniform and clean, adhering to the SiO$_2$. The Au is composed of many grains which are fused together to form the electrode due to the evaporation deposition. Although the contact edge is not straight, the shadow mask technique used has produced abrupt final grains which can be used as the edge of the contact and thus the junction location. A few grains of Au have leaked onto the MoS$_2$ region but STS line profiles are taken far from any such contamination. Topographies are both taken at set points of 2 V and 100 pA.

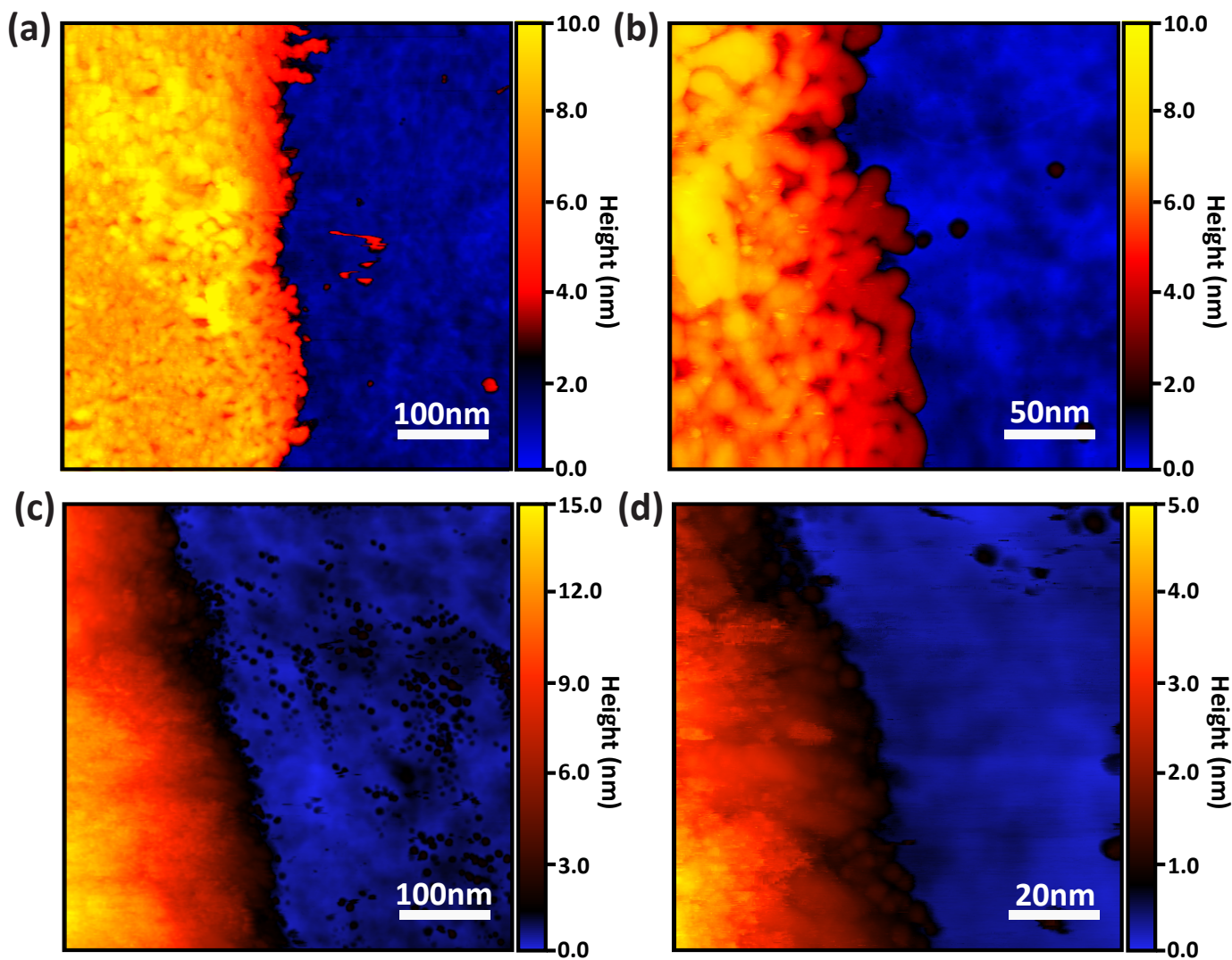

**Figure S1**. (**a-b**) STM topographic images of Au-MoS$_2$ junction taken at 2 V and 100 pA. The coalescence of grains can be seen on the Au contacts. We can use the final grains at the edge for a sharp Au-MoS$_2$ junction. (**c-d**) STM topographic images of Pd-MoS$_2$ junction taken at 2 V and 100 pA. The coalescence of grains can be seen on the Pd contacts. We can use the final grains at the edge for a sharp Pd-MoS$_2$ junction.

Figures S1c and S1d show STM topographies of the Pd-MoS$_2$ junction. As the Pd evaporation source was less stable than the Au, the electrode has slightly more leakage of metal particles onto the MoS$_2$ than the Au electrode. Albeit this, clean regions with sharp final grains far from leaked grains are plentiful. The STS line profiles are taken in regions far (tens of nm) from any metal leaked grains, thus ensuring that they do not influence the dI/dV heat maps. Topographies are both taken at set points of 2 V and 100 pA.

### S3. Polymer Transfer Details

In brief, a bulk graphite crystal was mechanically exfoliated onto a PPC-covered Si wafer. The PPC film was subsequently placed on a piece of PDMS on a glass slide, which was then inverted and the desired graphite flake positioned above the MoS2 film. After establishing contact between the two, the temperature was gradually increased to 90 C to melt the PPC and allow the PDMS to detach from the substrate. Finally, samples were soaked in acetone overnight and rinsed with isopropanol to remove any residual PPC.

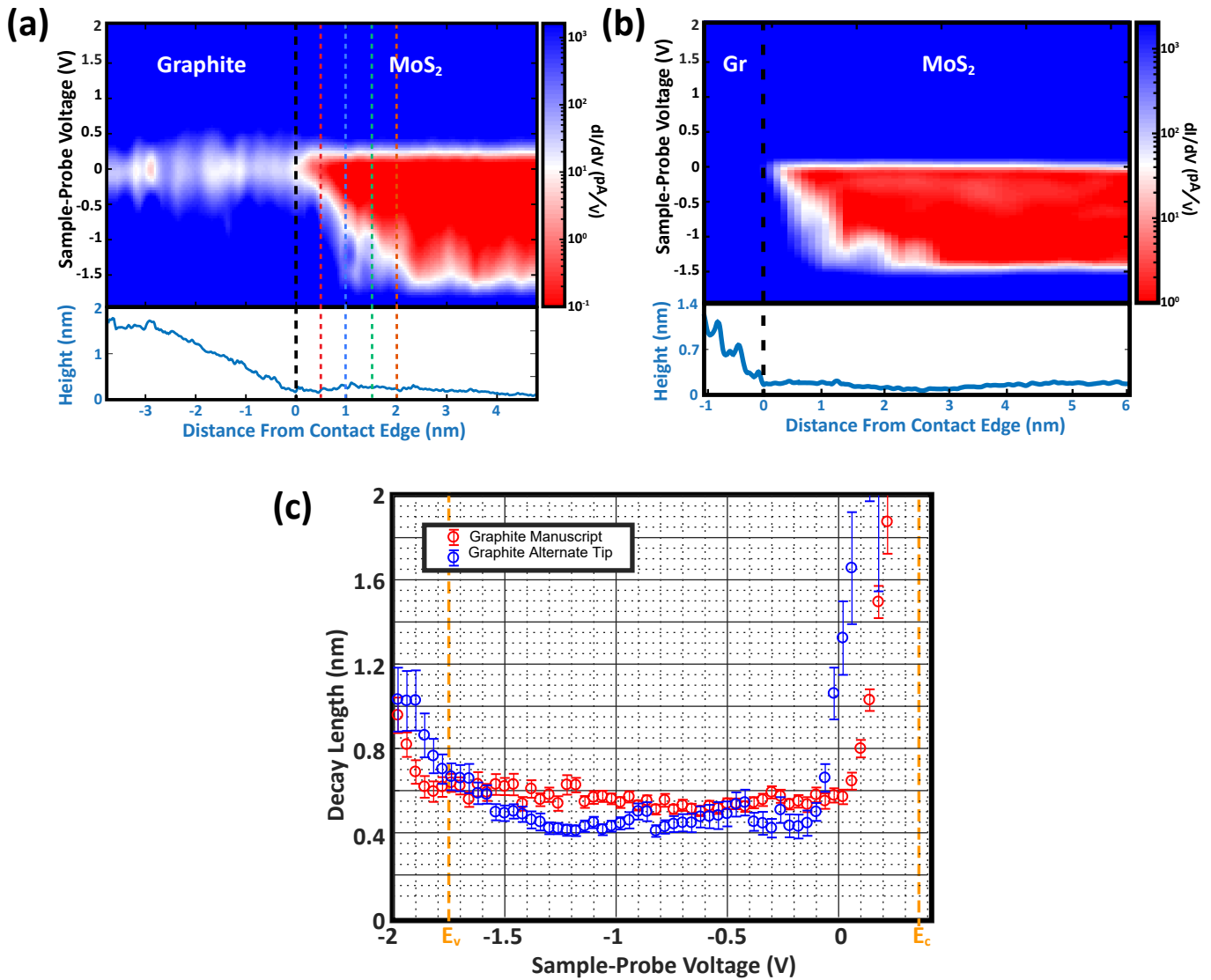

**Figure S2. (a-b)** Color maps of *short* (~9-nm) dI/dV line profiles taken across the same graphite-MoS$_2$ with two different tips, on two different regions. Below the color maps are simultaneously measured STM topography profiles revealing the approximate location of the contact edge. In both datasets, the color maps reveal a metallic L-DOS profile up until the edge of the electrode. Immediately to the right of the interface into the MoS$_2$ side, there is still no observed L-DOS gap. A L-DOS gap emerges at a finite distance into the MoS$_2$ side and grows for about 2 nm before reaching the far from junction magnitude. **(c)** A comparison of the decay lengths of the MIGS from the two datasets shown in figures S3 (a-b) showing reasonable agreement.

### S4. Shadow Mask Technique

A high quality razor blade is cut to a several mm$^2$ piece, keeping the razor edge clean and untouched. Razor blade piece is then positioned carefully onto samples using Kapton tape. Razor blade piece is positioned to sit at an angle such that the sharp razor blade edge is only about 5 μm from the sample surface.

## S5. Tip Comparison

As probe effects are a serious consideration, we have compared two data sets taken on the same junction, however with different tips and in different locations. Figure S2 (a-b) show a comparison of the Graphite-$MoS_2$ dI/dV 10 nm line cut provided in the manuscript, with an alternate dataset taken on the same Graphite-$MoS_2$ junction, however in a different area and with a different tip. Both show the distinct evolution of the gap. In Figure S2 (c), we have plotted the corresponding decay lengths of each of the two datasets, as in the analysis in Figure 4 (c) of the manuscript. The plot shows that the decay lengths of the two datasets are in reasonable agreement.